\documentclass[prx,aps,twocolumn,amsmath,amssymb,floatfix]{revtex4-1}

\pdfoutput=1
\usepackage{hyperref}
\usepackage{graphicx}
\usepackage[usenames]{color}
\usepackage{bm}

\def\cal#1{\mathcal{#1}}
\def\eqq#1{Eq.~(\ref{#1})}

\def\eq#1{(\ref{#1})}

\def\f#1{Fig.~\ref{#1}}

\def\s#1{Section~\ref{#1}}

\def\c#1{~\cite{#1}}

\def\cc#1{Ref.~\cite{#1}}

\def\r{\bm r}
\def\t{{\bm \tau}}
\def\ss{\bm S}
\def\e{{\rm e}}
\def\es{\epsilon_{\rm S}}
\def\d{{\rm d}}

\def\av#1{\langle #1 \rangle}

\def\o{{\cal O}}

\def\beq{\begin{equation}}
\def\eeq{\end{equation}}
\def\bea{\begin{eqnarray}}
\def\eea{\end{eqnarray}}

\begin{document}

\title{Deposition control of model glasses with surface-mediated orientational order}
\author{Stephen Whitelam}\email{swhitelam@lbl.gov}
\affiliation{Molecular Foundry, Lawrence Berkeley National Laboratory, 1 Cyclotron Road, Berkeley, CA 94720, USA}
\author{Peter Harrowell}\email{peter.harrowell@sydney.edu.au}
\affiliation{School of Chemistry, University of Sydney, Sydney New South Wales 2006 Australia}

\begin{abstract}
We introduce a minimal model of solid-forming anisotropic molecules that displays, in thermal equilibrium, surface orientational order without bulk orientational order.  The model reproduces the nonequilibrium behavior of recent experiments in that a bulk nonequilibrium structure grown by deposition contains regions of orientational order characteristic of the surface equilibrium. This order is deposited in general in a nonuniform way, because of the emergence of a growth-poisoning mechanism that causes equilibrated surfaces to grow slower than non-equilibrated surfaces. We use evolutionary methods to design oscillatory protocols able to grow nonequilibrium structures with uniform order, demonstrating the potential of protocol design for the fabrication of this class of materials.
\end{abstract}
\maketitle

\section{Introduction}
Deposition at a planar interface occurs in a broad range of material fabrication processes, including freezing from the melt, precipitation, sedimentation, electrodeposition, and vapor deposition\c{kirkpatrick1975crystal,matsui2000three,campbell2001science, yin2001fabrication,shimmin2006slow,ishikawa2010bone,kakran2010fabrication,edstrom2011electrodeposition,li2013precipitation}. When the rate of deposition exceeds the rate of molecular relaxation in the growing solid, the result is a nonequilibrium material, usually characterized by disorder that becomes more pronounced with increasing growth rate. Examples of this type of disorder include defect trapping in crystal growth\c{trapping1989}, and polycrystallinity formed during rapid sedimentation of colloidal suspensions\c{sediment2006}. 

There exist, however, interesting exceptions to the scenario in which ``nonequilibrium'' is synonymous with ``disorder'': nonequilibrium structures produced by deposition can be ordered, in some cases more so than their equilibrium counterparts. One of these exceptions was recently reported by Bishop et al.\c{bishop2019}, a vapor-deposited organic molecule that forms layered smectic structures despite the fact that no such phase exists in the equilibrium phase diagram. The authors proposed that the smectic structure instead reflects the nature of order at an equilibrated surface, and used NEXAFS to show that the equilibrium surface indeed exhibits smectic order\c{bishop2019}. The idea of deposition producing a form of bulk order inaccessible to the system at equilibrium was first identified by Hellman\c{hellman1994} in 1994. Hellman noted that vapor deposition typically favors the structure that minimizes the surface free energy, as opposed to the bulk free energy. The general proposal that vapor deposition, suitably tuned, can achieve a nonequilibrium structure characterized by an extensive accumulation of the equilibrium surface structure has been developed recently by Ediger and coworkers\c{ediger2017, ediger2019}.

The possibility of designing deposition protocols to exploit an accumulation of surface-induced organization is exciting because of the prominence of the surface in fabrication\c{kirkpatrick1975crystal,matsui2000three,campbell2001science, yin2001fabrication,shimmin2006slow,ishikawa2010bone,kakran2010fabrication,edstrom2011electrodeposition,li2013precipitation}, and because of potential for the design of nonequilibrium materials, which are susceptible to the various controls one can exert during fabrication. In this paper we demonstrate control of this nature using evolutionary learning of deposition protocols to make model glasses with defined orientational order. 

There have been a number of simulation studies of orientational order in vapor-deposited films using chemically accurate molecular models\c{lyubimov2015,dalal2015,walters2017}. Here we use a minimal model in order to identify the least detail required to reproduce key aspects of experimental phenomenology, and to demonstrate that fabrication control can be exerted using control parameters common to many experimental systems. We show that a model possessing a local orientational degree of freedom, and exhibiting surface orientational order different to that of the bulk, forms under a vapor-deposition protocol a nonequilibrium structure that contains order characteristic of the surface equilibrium, thereby reproducing the phenomenology of glasses made in experiment. 

In more detail, we consider a minimal 3D lattice model of anisotropic molecules that possess continuous orientational degrees of freedom. Model particles possess an energy of interaction that we intend to be broadly representative of solid-forming anisotropic molecules that display, in thermal equilibrium, surface orientational order without bulk orientational order. We perform dynamic simulations of the model in the grand-canonical ensemble, allowing particles to bind, unbind, and rotate when adjacent to free volume, but not in the bulk of a structure. Similar constraints are commonly used in models of glasses\c{fredrickson1984kinetic,butler1991origin}. These ingredients, which allow for the emergence of surface deposition\c{leonard2010}, are enough to reproduce the nonequilibrium behavior of recent experiments\c{swallen2007,swallen2009} in that the bulk of a grown structure contains regions of orientational order characteristic of the surface equilibrium.

 In the present model, under a simple slow-growth protocol, this order is deposited homogenously when the degree of equilibrium surface order is small, and heterogenously when it is large. The heterogeneity results from an emergent growth-poisoning mechanism that causes ordered surfaces to grow slower than non-ordered surfaces. The existence of this poisoning mechanism suggests the limitations of simple protocols to produce order on demand, and provides an opportunity to demonstrate the utility of protocol design for the fabrication of this class of materials: we show that evolutionary methods can learn a protocol of oscillating chemical potential and temperature in order to fabricate nonequilibrium structures with near-uniform order.

In \s{model} we introduce the model. In \s{eq} we study its equilibrium behavior, and show that the surface of a solid film exhibits orientational order while the bulk does not. In \s{growth} we show that slow growth results in structures whose bulk contains orientational order characteristic of the surface equilibrium, albeit distributed in general in a nonuniform way. In \s{learning} we show that evolutionary methods can learn oscillatory protocols that produce nonequilibrium structures containing uniform order. We conclude in \s{conc}.

\section{Model and simulation details}
\label{model}

We consider a three-dimensional cubic lattice of $N_0=x_0 \times y_0 \times z_0$ sites. Unless otherwise stated, all lengths are set to $30$ lattice units. Each site $i$ of the lattice can be vacant or occupied by a particle. Particles receive an energetic penalty of $\mu$ relative to vacancies. We will work in the grand-canonical ensemble, and so $\mu$ acts as a chemical potential that controls the concentration of particles in the notional bath to which the system is coupled; we will use it in subsequent sections to control the rate of growth of particle structures. 

Particles bear a continuous orientation vector $\ss$ that lives on the unit sphere. Two particles on adjacent lattice sites $i$ and $j$ experience a pairwise energy of interaction
\beq
\label{nrg}
E_{ij} = -\epsilon - \es \left[2-(\ss_i \cdot \r_{ij})^2 -(\ss_j \cdot \r_{ji})^2 \right].
\eeq
We work in units such that $k_{\rm B} =1$, and unless otherwise stated we set $\epsilon=1$ and $\es=4$. Physically, the combination $\epsilon + \es$ controls the energy scale of interparticle binding, while $\es$ controls the energy scale associated specifically with orientation. $\r_{ij}$ is the unit vector pointing from the center of lattice site $i$ to the center of lattice site $j$. The orientational component of $\eq{nrg}$ encourages neighboring particles to align their orientation vectors perpendicular to their separation vector, modeling the idea that elongated molecules receive an energetic reward by putting their long axes side-by-side. As we discuss later, existing vapor deposition experiments likely correspond to the case $\epsilon_S \lesssim \epsilon$. We shall consider this regime but focus on the case $\es \gg \epsilon$, in order to explore the regime of substantial orientational order (see \f{fig1}).

We apply periodic boundary conditions in the $x$- and $y$- directions. In the $z$-direction the plane $z=1$ is attractive to particles (with energy of interaction $-\epsilon$ per lattice site), and the plane $z=z_0+1$ is held vacant. 

We simulated the system using Monte Carlo methods\c{binder1986introduction}. At each timestep we pick at random a lattice site. If it is vacant we propose to place there a particle, randomly oriented. If the lattice site is occupied we attempt with probability $p_{\rm rot}$ to rotate the particle about a randomly-chosen axis by an angle chosen uniformly on $[-1/5,1/5]$ radians. With probability $1-p_{\rm rot}$ we attempt instead to make the site vacant. Unless otherwise stated we set $p_{\rm rot}=1/2$. We accepted proposed moves at lattice site $i$ with probability ${\cal C}_i\min\left(1, f \e^{-\Delta E/T}\right)$. Here $T$ is temperature; $\Delta E$ is the energy change under the proposed move; the factor $f$ is 1, $1/(1-p_{\rm rot})$, and $1-p_{\rm rot}$ for rotation, insertion, and deletion moves, respectively; and ${\cal C}_i$ is a kinetic constraint that is zero if all 6 nearest neighbors of lattice site $i$ are particles, and unity otherwise. 

This algorithm models physical vapor deposition from a notional bath. The dynamical rules satisfy detailed balance with respect to the model's energy function, meaning that sufficiently long simulations will sample the thermal equilibrium associated with the values of the parameters $\mu$, $\epsilon$, $\es$, and $T$. On finite timescales, however, there is no guarantee that we will reach equilibrium. In particular, the kinetic constraint makes relaxation with the bulk of a particle structure slow, which is physically appropriate for many materials\c{mehrer2007diffusion}. Starting from an empty simulation box we can assess how a particle structure grows, under conditions for which binding and unbinding and orientational relaxation can occur at a surface, but not within the bulk of a structure (except where mediated by vacancies). Our simulations will in principle achieve the bulk equilibration that is observed experimentally under very slow growth conditions, but we typically observe a separation of timescales such that kinetically trapped, orientationally ordered films take much longer to relax to bulk equilibrium than they do to grow. We measure time in units of $N_0$ attempted Monte Carlo moves. To assess equilibrium directly we start from a box of particles aligned in the $z$-direction and switch off the kinetic constraint, setting ${\cal C}_i = 1$ for all $i$.

\section{Equilibrium surface order without bulk order}
\label{eq}

\begin{figure}[] 
   \centering
\includegraphics[width=\linewidth]{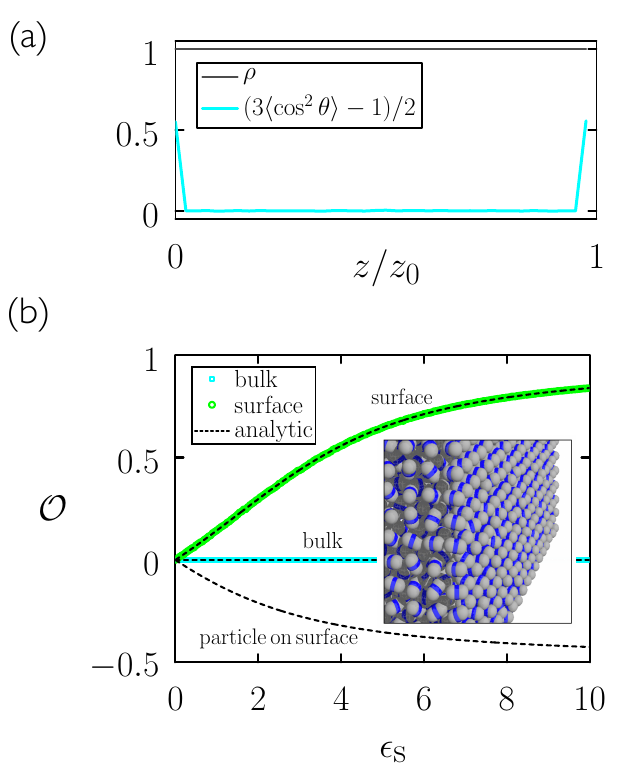} 
   \caption{\label{fig1} (a) Particle density $\rho$ (gray) and layer-averaged values of the nematic order parameter (cyan) of a structure in equilibrium [parameters: $\mu=0, T=1, \epsilon=1, \es=4, z_0=40$]. (b) Self-consistent mean-field theory (black dashed lines) and Monte Carlo simulations (colored symbols) show the nematic order parameter $\o$ to vanish in equilibrium in the bulk but not at a surface. Parameters are as panel (a), but with $\es$ varied. Errorbars are smaller than symbols. The snapshot shows the surface and a portion of the bulk for the case $\es=10$; the difference in orientational order between surface and bulk is visible to the eye.}
\end{figure}
We start by determining the behavior of the model in thermal equilibrium. In \f{fig1}(a) we show the equilibrium orientational order for a simulation box full of particles, obtained from equilibrium Monte Carlo simulations. We plot the layer-averaged value of the nematic order parameter 
\beq
\label{order}
\o \equiv \frac{1}{2} \left(3\av{\cos^2 \theta}-1\right)
\eeq
as a function of distance $z$ along the $z$-axis. Here $\theta$ is the angle between a particle's orientation vector and the box $z$-axis (the simulation box is anisotropic on account of the closed- and open boundaries in the $z$-direction). The angle brackets denote a thermal average over particles in each layer. As defined, ${\cal O}$ = 1 or $-1/2$ for perfect alignment parallel or perpendicular to the surface normal, respectively, and is zero for random orientations. It is clear from \f{fig1}(a) that, at equilibrium, there is no orientational order within the bulk of the structure, but there is normal alignment at the surface.

To understand this result we turn to self-consistent mean-field theory\c{binney1992theory,geng2009theory}. We impose the number of neighbors each particle possesses, and self-consistently calculate the resulting equilibrium orientational order. To this end we assume that a particle $i$ experiences an effective energy of interaction
\beq
\label{eff}
{\cal H}_{\rm eff}(\ss_i) = -\es\sum_j\left[2-(\ss_i \cdot \r_{ij})^2 -(\t \cdot \r_{ji})^2 \right],
\eeq
where the sum runs over the nearest neighbors of site $i$, and $\t  = \av{\ss_i}$ is the thermal average of the particle's orientation vector. Thermal averages are calculated self-consistently as $\av{(\cdot)} = {\rm Tr}\, (\cdot) \e^{-{\cal H}_{\rm eff}}/{\rm Tr} \, \e^{-{\cal H}_{\rm eff}}$, where ${\rm Tr} \equiv \int_{-1}^1 \d(\cos \theta) \int_0^{2 \pi} \d \phi$ with $\ss$ expressed in spherical polar coordinates. We assume temperature $T=1$. Inserting \eq{eff} into the average (and dropping site labels) we get
\beq 
 \label{av2}
\av{(\cdot)} = \frac{ \int_{-1}^1 \d(\cos \theta) \int_0^{2 \pi} \d \phi \,(\cdot) \e^{-\es \left[n_x S_x^2+n_y S_y^2+n_z S_z^2\right]}}{ \int_{-1}^1 \d(\cos \theta) \int_0^{2 \pi} \d \phi\, \e^{-\es\left[n_x S_x^2+n_y S_y^2+n_z S_z^2\right]}}.
\eeq
Here $n_x,n_y,n_z \in \{0,1,2\}$ are the number of neighbors of the particle in each direction, and the particle's orientation vector components are $S_x = \sin \theta \cos \phi$, $S_y = \sin \theta \sin \phi$, and $S_z = \cos \theta$. 

Using \eq{av2} we can calculate the value of $\av{\cos^2 \theta}$ and hence the nematic order parameter \eq{order} in thermal equilibrium. We show these results in \f{fig1}(b) as black dashed lines, together with the results of Monte Carlo simulations (colored symbols); analytic and simulation results agree.

In bulk we have $n_x=n_y=n_z=2$, and from \eq{av2} we get $\av{\cos^2 \theta} =1/3$. Hence $\o$ vanishes in bulk in thermal equilibrium for any value of $\es$ (see the line labeled ``bulk'' in \f{fig1}(b)). By contrast, at an exposed surface in the plane $z=$ constant we have $n_x=n_y=2$ and $n_z=1$, in which case we get from \eq{av2} that 
\beq
\label{surf}
\av{\cos^2 \theta} = \frac{\int_{-1}^1 \d x \, x^2 \e^{\es x^2}}{\int_{-1}^1 \d x \,\e^{\es x^2}}=\frac{\e^{\es }}{\sqrt{\pi \es } \, \text{erfi}\left(\sqrt{\es }\right)}-\frac{1}{2 \es },
\eeq
where erfi is the imaginary error function. \eqq{surf} and the resulting value of $\o$ increase with $\es$ (see the line labeled ``surface'' in \f{fig1}(b)), showing the surface to be orientationally ordered in equilibrium, with particles possessing a tendency to point parallel to the surface normal (i.e. perpendicular to the surface). Thus the presence or absence of a single neighboring particle profoundly changes the nature of equilibrium orientational order: an exposed surface is ordered but the bulk is not.

It is also useful, to understand the growth behavior of this model, to calculate the equilibrium order of a single particle on a free surface, for which $n_x=n_y=0$ and $n_z=1$. In this case the value of $\av{\cos^2 \theta}$ is given by \eq{surf} with the replacement $\es \to -\es$, and the resulting value of $\o$ is plotted in \f{fig1}(b) (see the line labeled ``particle on surface''). An isolated particle on a surface tends to align perpendicular to the surface normal, the more so as $\es$ increases.

\begin{figure}[] 
   \centering
\includegraphics[width=\linewidth]{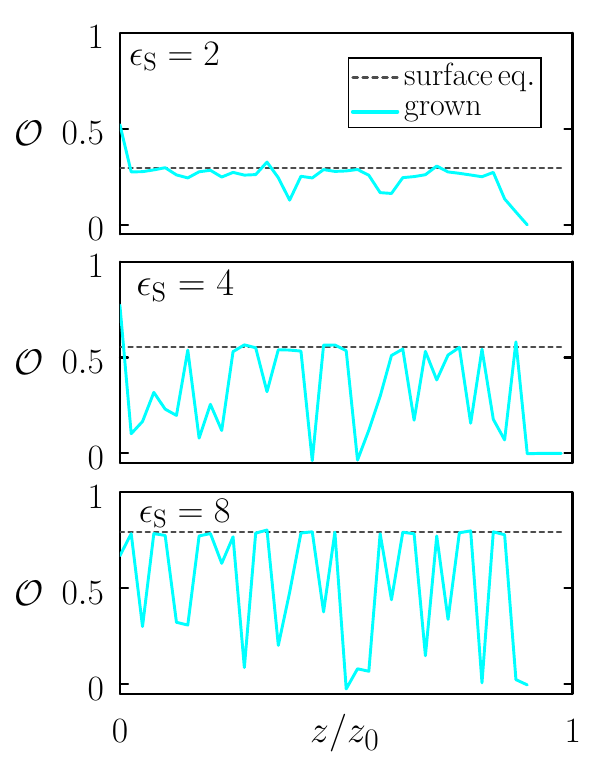} 
   \caption{\label{fig2} Order-parameter profile (cyan) of films deposited slowly using the values $\es = 2, 4$, and 8 of the anisotropic particle coupling. The associated values of the equilibrium surface order are given by the black dashed lines. The deposited structures are orientationally ordered, unlike their equilibrium counterparts (compare \f{fig1}(a)). As the degree of equilibrium surface order increases, the dynamical bulk order becomes irregular.} 
\end{figure}

\section{Grown structures are orientationally ordered but not, in general, uniform}
\label{growth}

In \f{fig2} we show the results of growth simulations carried out at three different values of the orientational coupling $\es$. In each case we carried out simulations at fixed temperature $T=1$ for several values of the chemical potential $\mu$, which controls the growth rate of the sample. The largest degree of bulk order is attained at the lowest accessible growth rates, and profiles along the growth direction of the resulting samples are shown in the figure. We indicate the value of the equilibrium surface order using a black dashed line.  In profiles, the order parameter ${\cal O}$ is given by \eq{order}, where now the average $\av{\cdot}$ is taken over the particles in each layer of a single sample (unlike the equilibrium averages of \f{fig1}(a), which are taken over many configurations). These simulation results reproduce a key feature of the experiments of Refs.\c{hellman1994,ediger2017,ediger2019}: with decreasing growth rate, the order parameter in the deposited film increases and approaches the value of the equilibrium surface. In the experiments of~\cc{bishop2020a}, decreasing the growth rate eventually results in a decrease in the film order as it achieves the unordered bulk equilibrium. This regime is accessible to our model but only at much longer simulation times than those considered here.

For the smallest coupling shown in \f{fig2}, $\es=2$, the order throughout the deposited film is essentially that of the equilibrium surface. This equivalence is consistent with the results of molecular dynamics simulations of deposition reported in~\cc{dalal2015}. In that paper, the orientational order of a film deposited at a temperature just below the bulk glass transition temperature $T_{\rm g}$ was the same as that of the equilibrium surface, ${\cal O} \approx 0.04$. The largest values of ${\cal O}$ reported in the experiments of that paper were approximately 0.2, similar to the order shown in the top panel of \f{fig2}. 

More generally, our model results suggest that achieving in bulk the value of the equilibrium surface order becomes more difficult as the surface becomes more ordered. Increasing the degree of order at the equilibrium surface results in large fluctuations in order of the deposited film. For the two larger couplings $\es$ shown in \f{fig2}, the deposited film harbors regions of equilibrium surface order, but distributed in a nonuniform way. The grown films exhibit large fluctuations corresponding to random oscillations between the value of the surface equilibrium and values close to zero, denying the deposited film the full order of the equilibrium surface. 

These structural fluctuations are of considerable interest because they limit the degree of order that can be captured during deposition. They also provide the structural sensitivity to growth conditions that create opportunities for machine-learning control, a topic we return to in \s{learning}. We first explore in more detail these structural fluctuations, whose origin lies in the different attachment kinetics associated with particles aligned parallel and perpendicular to the surface normal.

In \f{fig3}(a) we show (cyan symbols) the bulk order ${\cal O}$ obtained from growth simulations at various values of chemical potential $\mu$, for $\es=4$ (for which the value of the surface order parameter ${\cal O}$ in equilibrium is 0.556). Averages in this figure are taken over all non-surface particles in deposited films, and errorbars are computed by comparing as many films as could be produced in a fixed simulation time (the number of films varies considerably throughout the figure). As $\mu$ is increased and growth is slowed, the bulk order parameter approaches, but does not reach, the value of the surface equilibrium order parameter.
 
 \begin{figure*}[] 
   \centering
\includegraphics[width=\linewidth]{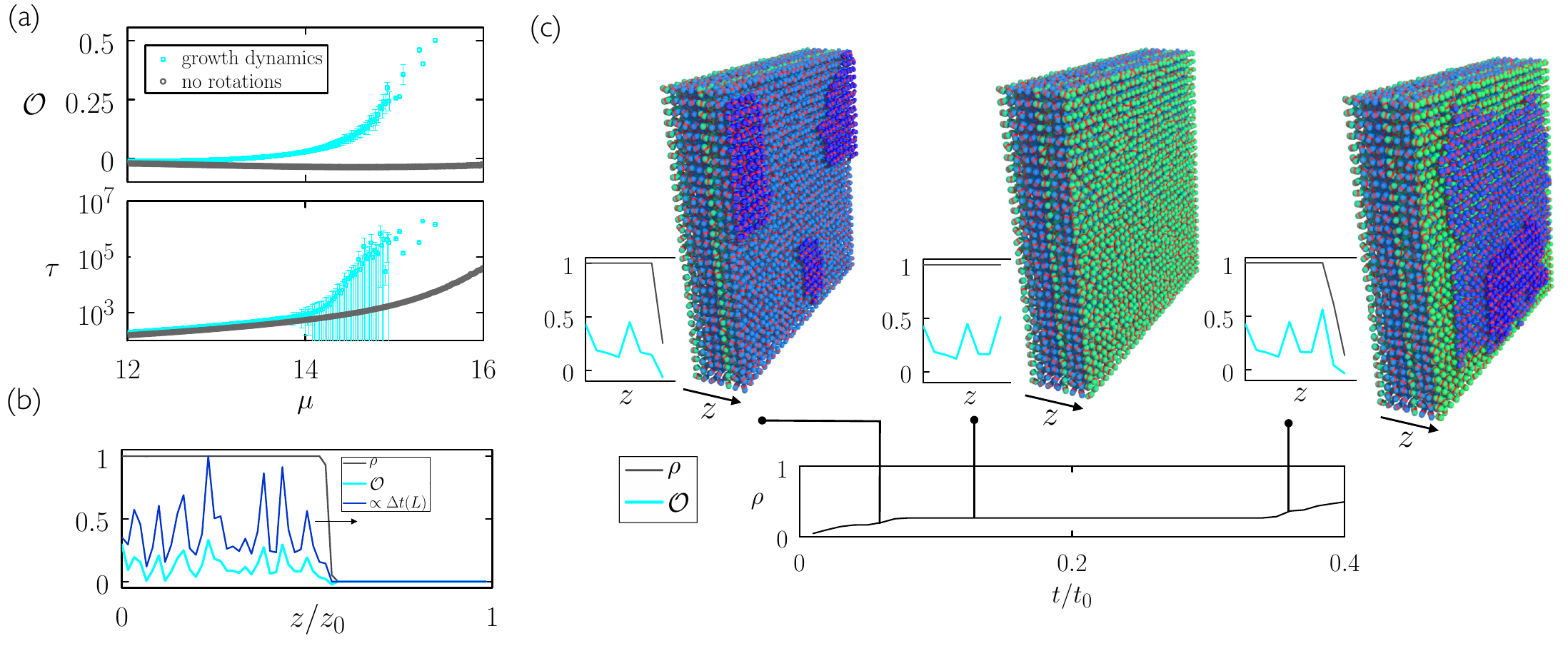} 
   \caption{\label{fig3} (a) Bulk orientational order (top) and growth time (bottom) for structures produced at various values of $\mu$ (cyan symbols). We show the same quantities for growth in which particle rotation is not allowed (gray symbols) [parameters: $T=1,\epsilon=1,\es=4$]. Errorbars indicate $\pm$ one standard deviation; the rightmost cyan points have no errorbars, because only one sample was obtained in the allotted time. (b) Order as a function of layer (cyan line) for a structure grown at $\mu =15$. We also indicate the average time for which particles in each layer were mobile (blue line). (c) Time series of growth. Particles in layers whose orientational order is close to values of 0.5 and 0 are colored green and blue, respectively.}
\end{figure*}

In the lower panel of \f{fig3}(a) we show the time taken to fill the simulation box as a function of $\mu$ (cyan symbols). We show the same quantity for simulations in which rotations are not permitted ($p_{\rm rot}=0$), so that particles retain the orientations they adopted during their deposition. Absent rotation, growth is faster, and can be observed at values of $\mu$ for which growth in the presence of rotation has arrested. In the presence of particle rotations, the tendency of particles in an exposed layer to align parallel to the surface normal slows growth, because the bond between an incoming particle and the surface is stronger if the surface particle points perpendicular to the surface normal. Surface equilibration, for sufficiently weak driving, is thus a form of self-poisoning\c{ungar2005effect,de2003principles,schilling2004self,higgs1994growth,ungar2000dilution,ungar2005effect,whitelam2016minimal}.

The tendency of equilibrated exposed surfaces to grow slower than non-equilibrated ones results in the structural fluctuations of order seen in \f{fig2}. In \f{fig3}(b) we show orientational order as a function of layer depth (cyan line) for a structure grown at one of the points shown in panel (a), faster than the structure grown in \f{fig2}. The order is again irregular. The blue line is proportional to the average time for which each particle in a layer was mobile, between its arrival and the time it first acquired 6 neighbors. The correlation between order and time mobile is clear.

In \f{fig3}(c) we show a time series of growth. The structure grows layer-by-layer, and the cause of the structural heterogeneity is as follows. As new particles are deposited on an existing layer (left-hand panel in \f{fig3}(c)), they experience a free-energetic impetus to point perpendicular to the surface normal (i.e. have small values of ${\cal O}$; see the line labeled ``particle on surface'' in \f{fig1}(b)). As the new layer grows laterally, particles acquire additional in-plane neighbors and experience a thermodynamic driving force to point parallel to the surface normal (i.e. have large values of ${\cal O}$; see the line labeled ``surface'' in \f{fig1}(b)). At this point a competition of timescales ensues: if the particles in the new layer achieve surface equilibrium then they become less sticky for new particles, and growth slows (middle panel in \f{fig3}(c)). If not, they become covered by a new layer and are rendered immobile (right-hand panel in \f{fig3}(c)). Crucially, whatever the rate of nucleation on an equilibrated surface, the rate of nucleation on a non-equilibrated surface is larger (\f{fig3}(a)); the result, at the lowest growth rates, is a series of layers whose orientational order alternates between that characteristic of surface equilibrium (i.e. large positive ${\cal O}$, indicative of particles pointing parallel to the surface normal), and that characteristic of the equilibrium of a single particle on a surface (i.e. negative values of ${\cal O}$).

\section{Learning protocols to achieve uniform order}
\label{learning}

\begin{figure}[] 
   \centering
\includegraphics[width=\linewidth]{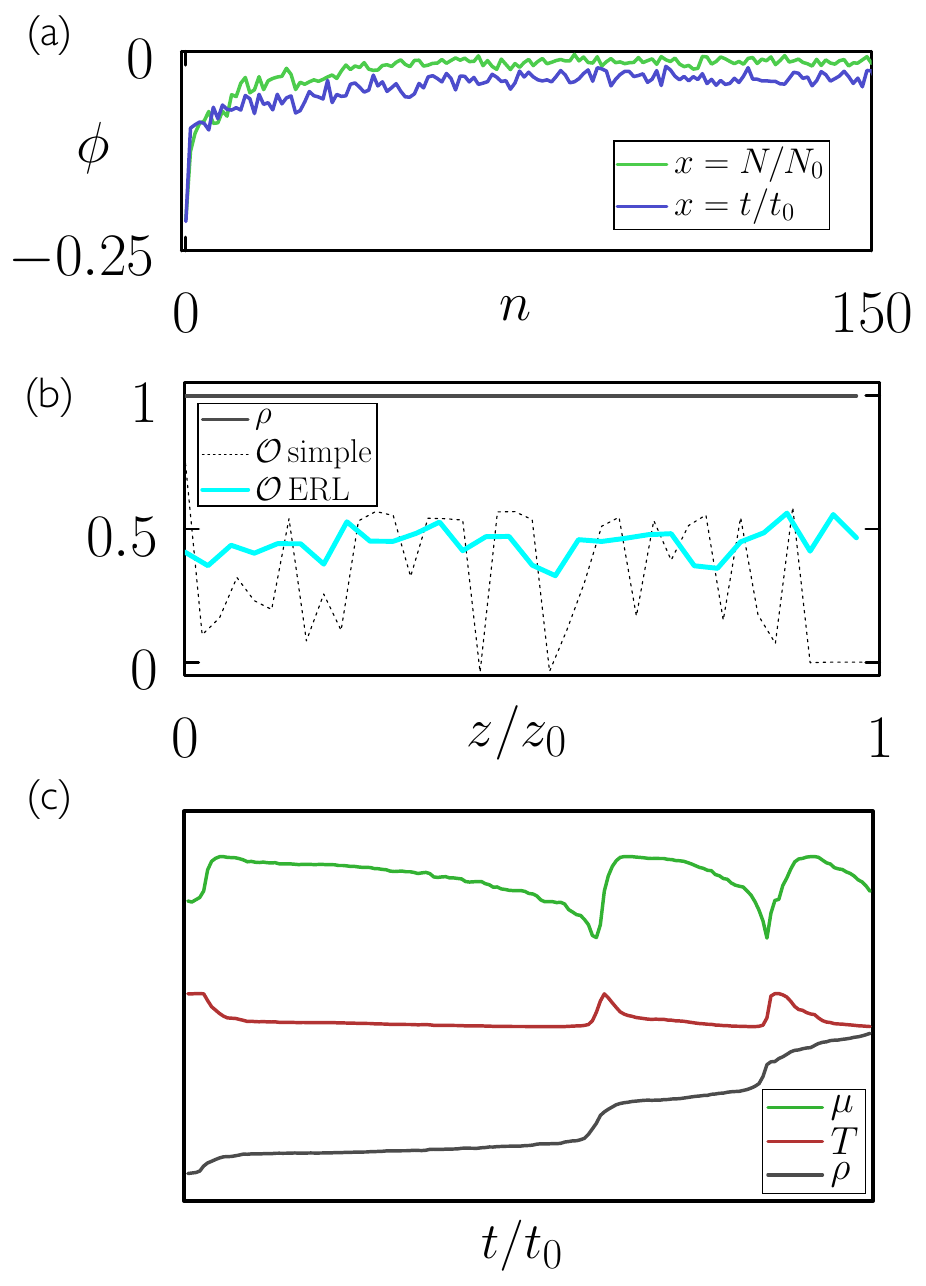} 
   \caption{\label{fig4} (a) Objective $\phi$ versus evolutionary time $n$ for two classes of learned protocol. (b) The order of a structure produced by the learned protocol (cyan line) is more uniform than that produced by a simple slow-growth protocol (black dashed line; this is the structure shown in the middle panel of \f{fig2}). (c) The protocol $\mu,T$ and number of deposited particles $\rho=N/N_0$ versus time for a segment of the trajectory used to produce the structure in panel (b). Lines have been scaled and offset in order to show the qualitative nature of the protocol.}
\end{figure}

We have highlighted the fact that grown structures are out of equilibrium, even under conditions of very slow growth. They incorporate the orientational order characteristic of the surface equilibrium, rather than the bulk equilibrium. The establishment of equilibrium surface order within the bulk of a nonequilibrium structure has been demonstrated experimentally. Dalal et al.\c{dalal2015} reported the deposition of three organic molecules that exhibit surface alignment ranging from weakly parallel to the surface normal ($0.1 < {\cal O} < 0.2$ for $0.9 < T/T_{\rm g} < 1$) to near-perfect alignment perpendicular to the surface normal ($-0.4 < {\cal O} < -0.3$ for $T/T_{\rm g} \ll 0.9$); note that ${\cal O}$ = 1 or $-1/2$ for perfect parallel and perpendicular surface-normal alignment, respectively. Since the rate of surface relaxation decreases with decreasing temperature, the transition from positive to negative $\cal O$ on cooling corresponds to a change in structural selection from (surface) thermodynamic to kinetic control.

Our model also shows that while films incorporate the orientational order characteristic of the surface equilibrium, they do so imperfectly, the more so as the degree of equilibrium surface order increases. This imperfection results from a difference in growth rates between surfaces of different degrees of order. The strategy of waiting as long as possible does not produce uniform order: the growth rate of surface-equilibrated layers is always less than that of their nonequilibrated counterparts [\f{fig3}(a)], and grown structures are heterogenous at the lowest growth rates we can achieve. Given the generic nature of the model and its success in capturing key features of experiments, this imperfection suggests limits to the order that can be captured by simple growth protocols.

In order to produce structures of uniform order, we turn to protocol design\c{miskin2016turning,tang2016optimal,whitelam2020learning,whitelam2020neuroevolutionary}. The strategy used thus far was to impose a value of chemical potential $\mu$ and temperature $T$ and wait for growth to happen. In this section we instead use evolutionary methods\c{GA,salimans2017evolution,Guber,whitelam2020learning,whitelam2020neuroevolutionary} to learn a protocol in order to achieve a desired structure. To do so we fix the time of simulation (we choose a value of $t_0=10^4$ Monte Carlo sweeps) and select an objective function 
\beq
\phi = -(z_0-2)^{-1}\sum_{j=2}^{z_0-1}({\cal O}_j-1/2)^2.
\eeq
Here ${\cal O}_j$ is the mean value of the orientational order parameter for particles in layer $j$ of a structure. The quantity $\phi$ is maximized if a grown structure (excluding the first and last layers) has uniform orientational order $1/2$.

For the protocol we use the ansatz 
\bea
\label{ansatz}
1/T &=& a_0 + a_2\sin\left(2 \pi a_4 x +a_6\right) \nonumber \\
\mu &=& a_1 + a_3\sin\left(2 \pi a_5 x +a_7\right),
\eea
containing parameters $a_i$, reasoning that periodically varying $T$ and $\mu$ is an appropriate choice for a structure deposited layer-by-layer. We choose the quantity $x$ in \eq{ansatz} to be either the normalized elapsed time $t/t_0$ or the normalized number of deposited particles $N/N_0$. 

In order to learn the parameters $a_i$ that maximize the objective $\phi$, we follow the evolutionary reinforcement learning (ERL) procedure of Refs.\c{whitelam2020learning,whitelam2020neuroevolutionary}. We created 100 different random initializations of \eq{ansatz}  (initial parameter values were $a_0=1,a_1=15,a_4=a_5=z_0=30,a_2=a_3=a_6=a_7={\cal N(0,0.01)}$), and used each within an independent simulation of fixed time $t_0$. We identified the 10 simulations with the largest values of $\phi$, evaluated at the end of each simulation. We then created a new population of 100 protocols by drawing randomly with replacement from the set of 10 and, for each, adding independent Gaussian random numbers ${\cal N}(0,0.01)$ to each parameter $a_i$. We ran a new simulation using each of these 100 new protocols, identified the 10 protocols with the largest values of $\phi$, and continued iteratively.

In \f{fig4}(a) we show the value of the objective $\phi$ after $n$ iterations (or ``generations'') of this evolutionary procedure. The protocol class with $x=N/N_0$ is better than that with $x=t/t_0$. In the former case the objective is essentially satisfied by the protocol produced after about 100 iterations. Were it not, one could consider more complicated protocols, such as those expressed by neural networks\c{whitelam2020learning,whitelam2020neuroevolutionary} (\eq{ansatz} can be regarded as a single-layer neural network with two hidden nodes and sine activation functions; increasing the number of hidden nodes or the depth of the network would allow it to express more general functions).

In \f{fig4}(b) we show the layer-by-layer order of one of the structures produced by the learning procedure (cyan line). The structure shows small variations of order by layer, but is close to being uniform. By contrast, the black dashed line shows the order of the structure generated at the lowest accessible growth rate from \f{fig3}(a), which is very far from being uniform. It is also of note that the learned protocol was restricted to an operation time of $t=10^4$ Monte Carlo cycles, while the slow-growth protocol took 238 times longer to produce a structure of the same size. This system is therefore another example in which a rapid, far-from-equilibrium protocol produces a more ordered structure than a simple ``wait as long as we can'' strategy\c{whitelam2018strong,whitelam2020learning}.

In \f{fig4}(c) we show the protocol $T$, $\mu$, and the number of deposited particles $\rho$ for a segment of the trajectory that produced the structure of panel (b). The protocol learned by the evolutionary algorithm heats each layer as it begins to appear, simultaneously reducing the thermodynamic driving force for growth, and reverses these trends as the layer nears completion.

\section{Conclusions}
\label{conc}

In this paper we have explored a model of orientational order in vapor-deposited films using computer simulations. Our lattice model, while simple, reproduces experimental results showing the important influence of the equilibrium surface structure in the deposited film, and highlights the role of deposition rate and interfacial relaxation times in selecting the orientational order of the nonequilibrium material. When the surface stabilization of molecular alignment is weak, we find that vapor deposition can achieve a homogeneous film of maximal orientational order via a simple search in the space of deposition temperature and rate. 

As the strength of surface alignment increases, however, so too does the magnitude of fluctuations in the orientational order parameter throughout the deposited film. These fluctuations are sensitive to details of the deposition protocol, and we have shown that adjustment of the deposition conditions to control the structural outcome can be achieved by protocol learning. Selecting orientational homogeneity as the objective, we have demonstrated using evolutionary reinforcement learning that a time-dependent protocol of temperature and chemical potential can achieve near homogeneous alignment within the deposited film. The degree of uniformity obtained is considerably greater than that afforded by fixed values of $T$ and $\mu$. 

It would be interesting to test experimentally whether our prediction of increased orientational order in the equilibrium interface results in an increase in the fluctuations of order within the deposited film. (Strategies for increasing surface order might include using glass-forming molecules with more pronounced anisotropy or that have a tendency to exhibit orientational order in bulk.) Whether or not the fluctuations we observe can be reproduced in experiment may depend on the choice of ensemble. In this paper we use an ensemble in which the chemical potential is imposed and the instantaneous rate of deposition can fluctuate. It is these fluctuations that contribute significantly to the fluctuations of orientational order. Experimental vapor deposition is sometimes carried out at very low gas phase densities and low effective temperatures. Under these conditions, deposition is effectively irreversible and the growth rate is approximately constant, fixed by the flux from the source. We leave to future work the clarification of the dependence of order fluctuations on the choice of growth ensemble.  

The goal of this paper is to introduce a minimal model of vapor deposition in a molecular system whose surface and bulk equilibrium order differ, and to examine the potential for exploiting protocol learning as a means of accessing the considerable structural variations of disordered materials through fabrication. Our results provide clear support for this program with the caveat that fabrication control is most usefully exercised via variables that fluctuate significantly under the fabrication conditions. Not all materials and fabrication situations will meet this requirement. Future work will be directed at understanding and classifying the types of structural variables that can facilitate the sought-after level of machine control.

\section{Acknowledgments}

We gratefully acknowledge Mark Ediger for valuable comments on the paper. This work was performed as part of a user project at the Molecular Foundry, Lawrence Berkeley National Laboratory, supported by the Office of Science, Office of Basic Energy Sciences, of the U.S. Department of Energy under Contract No. DE-AC02--05CH11231. P.H.acknowledges support from the Australian Research Council.


%

\end{document}